\documentclass[journal]{IEEEtran}

\usepackage{ifthen}
\newboolean{showcomments}
\setboolean{showcomments}{true}
\newcommand{\andrey}[1]{\ifthenelse{\boolean{showcomments}}
	{ \textcolor{blue}{(Andrey says:  #1)}}{}}
	
\usepackage{booktabs} 
\usepackage{multirow} 
\usepackage{multicol}
\usepackage{balance}
\usepackage{cite}
\usepackage[switch]{lineno} 
\usepackage{enumerate}
\usepackage{dsfont}
\usepackage{algorithm}
\usepackage[noend]{algpseudocode}
\usepackage{colortbl}

\makeatletter
\long\def\@makecaption#1#2{\ifx\@captype\@IEEEtablestring%
\footnotesize\begin{center}{\normalfont\footnotesize #1}\\
{\normalfont\footnotesize\scshape #2}\end{center}%
\@IEEEtablecaptionsepspace
\else
\@IEEEfigurecaptionsepspace
\setbox\@tempboxa\hbox{\normalfont\footnotesize {#1.}~~ #2}%
\ifdim \wd\@tempboxa >\hsize%
\setbox\@tempboxa\hbox{\normalfont\footnotesize {#1.}~~ }%
\parbox[t]{\hsize}{\normalfont\footnotesize \noindent\unhbox\@tempboxa#2}%
\else
\hbox to\hsize{\normalfont\footnotesize\hfil\box\@tempboxa\hfil}\fi\fi}
\makeatother

\usepackage{fancyhdr}
\makeatletter

\newif\ifdraft

\ifdraft
\makeatother
\fi

\usepackage{hyperref}
\hypersetup{
  unicode=false,          
  pdftoolbar=true,        
  pdfmenubar=true,        
  pdffitwindow=false,     
  pdfstartview={FitH},    
  pdftitle={My title},    
  pdfsubject={Subject},   
  pdfcreator={Creator},   
  pdfproducer={Producer}, 
  pdfkeywords={keyword1} {key2} {key3}, 
  pdfnewwindow=true,      
  colorlinks=true,       
  linkcolor=black,          
  citecolor=black,        
  filecolor=black,      
  urlcolor=black           
}

\ifCLASSINFOpdf
\usepackage[pdftex]{graphicx}
\graphicspath{{./}}
\DeclareGraphicsExtensions{.pdf,.jpeg,.eps,.png}
\else
\fi

\usepackage[cmex10]{amsmath}
\usepackage{amssymb}

\usepackage[caption=false,font=footnotesize]{subfig}

\newcommand{\tilX}{\tilde{X}}
\newcommand{\tilY}{\tilde{Y}}
\newcommand{\tilI}{\tilde{I}}

\def\BState{\State\hskip-\ALG@thistlm}

\begin{document}

\title{Aggregating Privacy-Conscious Distributed Energy Resources for Grid Service Provision}
\renewcommand{\theenumi}{\alph{enumi}}
\author{
  
  \IEEEauthorblockN{Jun-Xing~Chin,~\IEEEmembership{Member,~IEEE,}
  Andrey Bernstein,~\IEEEmembership{Member,~IEEE,}
  \\and~Gabriela~Hug,~\IEEEmembership{Senior Member,~IEEE}}%
  
  \thanks{This work was supported in part by the \emph{Swiss National Science Foundation} for the COPES project under the CHIST-ERA Resilient Trustworthy Cyber-Physical Systems (RTCPS) initiative.}%
  
  \thanks{J.X. Chin, and G. Hug are with the Power Systems Laboratory, ETH Zurich, 8092 Zurich, Switzerland. Email: \{chin $|$ hug\}@eeh.ee.ethz.ch. A. Bernstein is with the National Renewable Energy Laboratory, Golden, CO 80401, USA. Email: Andrey.Bernstein@nrel.gov}%
}

\markboth{Preprint}%
{Shell \MakeLowercase{\textit{J. Chin et al.}}: Aggregating Privacy-Conscious Distributed Energy Resources for Grid Service Provision}

\maketitle
\IEEEpeerreviewmaketitle

\begin{abstract}
The increasing adoption of advanced metering infrastructure has led to growing concerns regarding privacy risks stemming from the high resolution measurements. This has given rise to privacy protection techniques that physically alter the consumer's energy load profile, masking private information by using localised devices, e.g. batteries or flexible loads. Meanwhile, there has also been increasing interest in aggregating the distributed energy resources (DERs) of residential consumers to provide services to the grid. In this paper, we propose an online distributed algorithm to aggregate the DERs of privacy-conscious consumers to provide services to the grid, whilst preserving their privacy. Results show that the optimisation solution from the distributed method converges to one close to the optimum computed using an ideal centralised solution method, balancing between grid service provision, consumer preferences and privacy protection. More importantly, the distributed method preserves consumer privacy, and does not require high-bandwidth two-way communications infrastructure.
\end{abstract}


\section{Introduction}

In recent years, the adoption rate of advanced metering infrastructure (AMI) using smart meters (SMs) has risen steadily across the globe as part of grid modernisation efforts. As of January 2017, 52\% of the 150 million electricity consumers in the US have AMI \cite{EIA2017}, while in Europe, 13 member states are expected to have AMI adoption rates of over 95\% by 2020 \cite{EC2014}. In Switzerland, 80\% of all electricity meters are to be replaced with SMs by 2027 \cite{BFE2017}. AMI provides high-frequency energy consumption measurements to utility operators, allowing for data-driven grid management and planning techniques that promise to improve grid efficiency and transparency. However, this data also entails serious privacy risks for consumers, as it reveals their detailed electricity consumption profiles. Recent studies have shown that potential illnesses, religious practices, socio-demographic profile, and even appliances used can be inferred from AMI data through data analytics and non-intrusive load monitoring techniques \cite{Hargreaves2010,Molina-Markham2010,McDaniel2009,Becker2018,Wang2019}.

\subsection{Smart Meter Consumer Privacy}
These risks and recent developments in consumer privacy protection laws such as the European Union's General Data Protection Regulation \cite{EUGDPR16}, have spurred the development of privacy-enhancing methods for consumers with AMI, which can be split into two categories: \textit{smart meter data manipulation} (SMDM) and \textit{user demand shaping} (UDS) \cite{Giaconi2018a}. SMDM involves pre-processing the AMI data before it is reported, e.g., data aggregation, data anonymisation, and data obfuscation. UDS, on the other hand, entails physically shaping the consumer demand such that the grid-visible load, \emph{i.e.}, the \textit{grid load} no longer reveals private information present in the actual privacy-sensitive \textit{consumer load}. This is achieved by using behind-the-meter resources such as energy storage devices, flexible loads, and distributed energy sources. While the former may be cheaper to implement, they may impact the utility of the AMI data due to the distortion in the meter readings or may require trusted third parties, e.g, as proposed in \cite{Efthymiou2010,Rottondi2012}. Moreover, as they do not tackle the issue on the physical level, \emph{i.e.}, the actual energy flow, it might be possible to still decipher the actual consumption depending on the protection used \cite{Jaruwek2011}. 

One of the first UDS schemes is described in \cite{Kalogridis2010}, where the authors implement a best effort scheme to keep the grid load constant. However, this has been shown to leak information whenever there is a change in grid load \cite{Yang2012}, and has since been followed up by more complex schemes such as \cite{Zhang2017,Chin2017,Sun2018,Giaconi2018b}. In \cite{Zhang2017}, the authors propose a differential-privacy based protection scheme using an ideal battery to mask the on/off status of appliances while being cost-friendly. Using a model-distribution predictive control (MDPC) scheme that balances between minimising energy cost and a proxy for privacy loss, the authors in \cite{Chin2017} show that a home energy controller can be designed to directly minimise an approximate of mutual information between the grid and consumer loads. And in \cite{Sun2018}, the authors propose Q-learning based privacy-enhancing control policies using electric vehicles (EVs), flexible thermal loads, and energy storage devices to overcome limitations in modelling consumer load statistics. The control policies are tested on simulated load profiles with an ideal battery and a linearised thermal load model, and show that reasonable privacy-cost trade-off can be achieved by combining a small battery with EVs and an air conditioning device. The authors in \cite{Giaconi2018b} derived fundamental bounds on mutual information privacy for consumers with renewable energy sources (RES), both with and without an infinite battery, and proposed a sub-optimal privacy-enhancing scheme for realistic cases (finite battery) based on stochastic gradient descent. Note that in the absence of a battery, privacy can be enhanced through the curtailment of the available RES production.

\subsection{Residential Demand Side Aggregation and UDS Schemes}
Meanwhile, the increasing availability of behind-the-meter resources, a.k.a. distributed energy resources (DERs), coupled with the roll-out of smart grid communications infrastructure has spurred the development of demand-side management for smaller loads. Residential consumers, which have traditionally been neglected due to their size, are being aggregated in order to provide services to the grid. Residential demand side aggregation (RDSA) schemes can be divided into two main classes: direct load control, and incentive (signal) based schemes. The authors in \cite{Chapman2016} provide an overview of incentive based RDSA literature, and propose a multi-agent non-cooperative game framework for integrating RDSA schemes with home energy management systems (HEMSs). Nonetheless, limitations in communications infrastructure remain a challenge for most RDSA schemes\cite{Rajabi2017}. One possible solution is the broadcast of a common signal as suggested in \cite{Callaway2011}, though the design of the signal is still an active field of research; see, e.g., \cite{Bernstein2019} and pertinent references therein. 

UDS privacy protection methods, by their nature, lend themselves well to being a part of an RDSA scheme given their inherent flexibility to alter grid load. Moreover, it is intuitive that privacy-conscious consumers with UDS protection be considered potential DERs that can be aggregated to provide services to the grid. However, there are only few works on the design of privacy-centric HEMSs for RDSA schemes, with most works focusing solely on the auction activation, or (and) accounting mechanisms, e.g., \cite{Li2014, Gong2016, Balli2018}. These works employ different cryptographic techniques in the RDSA mechanism in order ensure privacy while being able to attribute rewards to demand response participants, but do not provide designs for the automated HEMSs. 

\subsection{Contributions and Outline}
In this paper, we design a HEMS for RDSA that also considers the consumers' preferences and privacy loss due to high-frequency SM measurements (SM privacy), without the need for pervasive real-time AMI metering, a trusted third-party, nor two-way high-bandwidth communications infrastructure to each consumer. Using an online projected gradient descent approach based on the framework in \cite{Dall'Anese2017} and the SM privacy protection scheme in \cite{Chin2017}, the proposed RDSA scheme preserves the SM privacy of consumers through UDS and omits the need for real-time information from each consumer. 

It is important to note that the allocation of rewards in RDSA schemes that do not directly control consumer loads typically rely on benchmarking metered consumption against historical records. However, in private-by-design RDSA algorithms, save for a totally flat profile, individual benchmark profiles should not be discernible from measured grid-visible consumer load profiles. Otherwise, private information could be inferred from repeated disclosures of a private, but static grid-visible load profile. Hence, rewards for participation in a private-by-design RDSA scheme in the absence of any trusted party can only be allocated based on the individual consumer's resource commitment. While this could result in a free-loader problem, such as those explored in game theoretic works, the resolution of this is outside the scope of this paper.

The rest of this paper is organised as follows: Section \ref{GenProb} details the problem considered, Section \ref{solutionmethod} introduces the distributed solution method, Section \ref{imp} presents and discusses simulation results and Section \ref{Conclusion} concludes the paper, and presents an outlook for future work.
  
\section{Problem Formulation and System Description}
\label{GenProb}
\begin{figure}
    \centering
    \subfloat[RDSA Aggregator System]{
    \includegraphics[trim=0cm 0.2cm 0cm 0.1cm, clip=true, width=0.85\columnwidth]{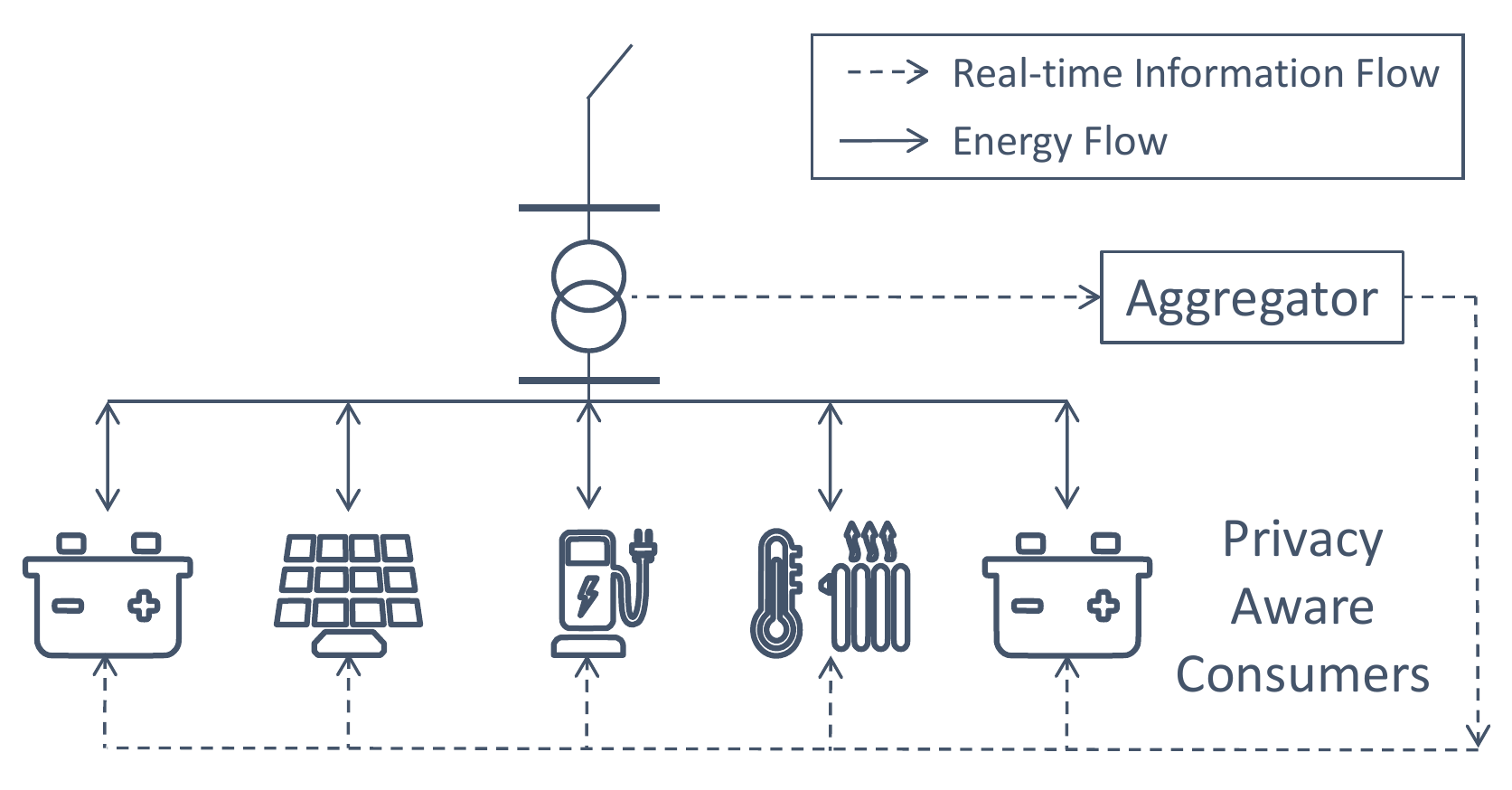}
    \label{fig:AggregatorSys}
    }\\ 
    \centering
    \subfloat[Setup at Consumer Household]{
    \includegraphics[trim=0cm 0.2cm 0cm 0.2cm, clip=true, width=0.85\columnwidth]{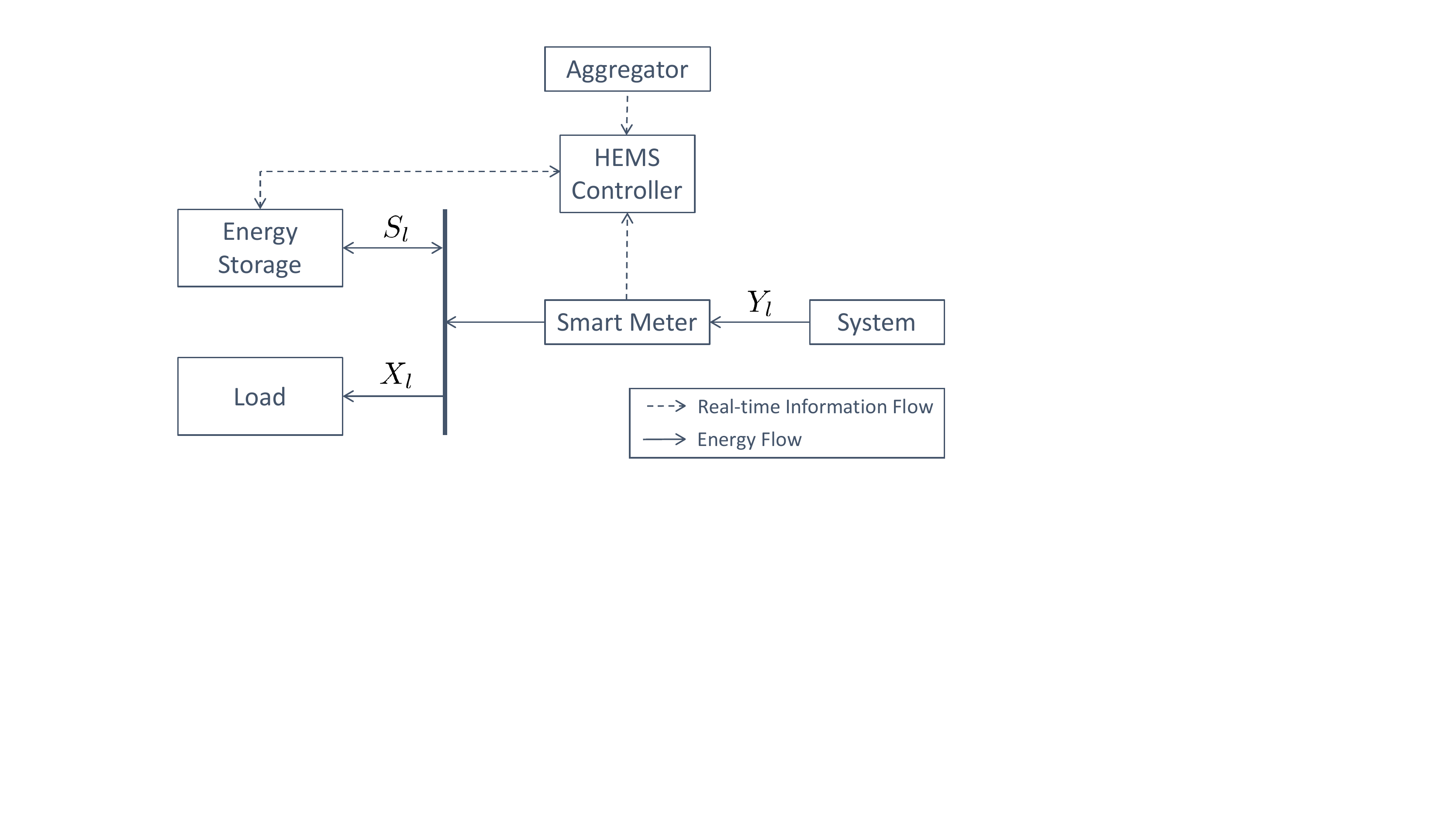}
    \label{fig:HomeSys}
    }
    \caption{System setup at consumer households and the aggregator system}
\end{figure}

We consider the problem of a HEMS that is part of an RDSA scheme and is required to consider the consumer's preferences, and also privacy-loss stemming from high-frequency smart meter measurements. Each consumer household consists of a DER, an HEMS controller, and a smart meter that is able to provide high-frequency measurements \textit{locally}, but unable or unwilling to provide \textit{real-time} high-frequency remote measurements to the utility provider or aggregator. The aggregator could be a distribution system operator or a third-party energy services provider that has the capability to broadcast high-speed uni-directional signals to each consumer, and is able to measure the real-time aggregated energy consumption of its consumers, e.g., at the sub-station or transformer. The general aggregator system setup is illustrated in Fig. \ref{fig:AggregatorSys}. On the other hand, the adversary considered in the privacy problem is assumed to be any entity (including an untrusted utility provider) that has access to the high-frequency smart meter measurements, \emph{i.e.}, there are no assumed trustworthy parties to which unprotected data is directly disclosed to. An ideal solution to the considered problem would directly disclose only the following data to third-parties:
\begin{enumerate}[i.]
    \item time-delayed high-frequency SM measurements from individual consumers that are privacy-protected; 
    \item real-time high-frequency aggregated consumer consumption measurements at points-of-common-coupling.
\end{enumerate}

Additionally, we consider the case of the DER being a battery in this paper, but the proposed method can easily be extended to incorporate other DERs with convex models. Fig. \ref{fig:HomeSys} illustrates the system considered at each consumer household. The problem can be framed as solving the following optimisation program:
\begin{alignat}{2}
\begin{split}\label{eq:objective}
& \underset{y_l}{\mbox{minimise}} \quad \sum_{l=1}^{N} \left\{ \Lambda(y_{l}) + \mu_l \Phi(y_l) \right\} + \rho \Gamma(\textbf{y}) \\
& \mbox{subject to} \quad y_l \in \mathcal{F}_l, \quad l = 1, \ldots, N,
\end{split}
\end{alignat}
where $N$ is the number of consumers in the aggregation, $y_{l}$ and $\mu_l$ are the grid load and price of privacy loss for consumer $l$, respectively,  $\Lambda(y_{l})$ is the consumer's utility (preference) function, $\Phi(y_l)$ is a measure of real time privacy-loss, $\rho$ is the coefficient for grid service provision, $\Gamma(\textbf{y})$ gives a measure of the grid service provided (e.g., target load or ancillary service tracking signal), and $\textbf{y} := [y_1,y_2,\cdots,y_N]^\mathsf{T}$ is a vector of consumer grid loads. The set $\mathcal{F}_l$, defined by the constraints:
\begin{alignat}{1}
&0 \leq ~s^+_l \leq s^{+,max}_l \label{eq:OriConStart}\\
&0 \leq ~s^-_l \leq s^{-,max}_l \\
&0 \leq ~e_l + \Delta t(\eta_l s^+_l - \frac{1}{\eta_l}s^-_l)  \leq e_l^{cap}\\
&y_l = x_l + s^{+}_l - s^{-}_l \label{eq:powerbalance}\\
&y^{min}_l \leq ~y_l \leq y^{max}_l \label{eq:OriConEnd}
\end{alignat}
enforces the system constraints for consumer $l$. Here, $s^+_l$, $s^-_l$ denote the battery's charging and discharging power; $s^{-,max}_l$, $s^{+,max}_l$ denote the battery's maximum discharging and charging power rating;  $\eta_l$ denotes the battery's charge/discharge efficiency; $e_l$ is the battery state of charge; $e_l^{cap}$ is the battery capacity;  $y^{min}_l$ and $y^{max}_l$ are the minimum and maximum allowable grid loads; $x_l$ denotes the instantaneous consumer load; and $\Delta t$ is the time interval between each control action. 

For the rest of the paper, $A_{l}$ denotes a random variable, $a_{l}$ denotes its realisation, $\log$ is the base-2 logarithm, $A^{t}$ denotes the sequence ($A_1, A_2, \cdots, A_t$), $\mathcal{A}$ is the range space for variable $A$, and $p_A(a)$ is the probability of $A=a$. The functions $\Lambda(y_{l}) ~, \Phi(y_l)$ and $\Gamma(\textbf{y})$ are described in the following. 

\subsection{Consumer Preferences}
The function $\Lambda(y_l)$ can be any convex utility function that reflects the consumer's preferences. For simplicity, we consider a function that penalises deviations from the consumer's day-ahead grid-load schedule, given by 
\begin{equation}\label{eq:preference}
    \Lambda(y_l) := \|y_l - y^\textit{ref}_l\|^2_2~,
\end{equation}
where $y^\textit{ref}_l$ is the day-ahead planned consumption. This schedule can include the consumer's preferences on a grid visible load profile, energy cost optimisation and battery state-of-charge requirements.



\subsection{Privacy Loss Proxy}
There has yet to be a consensus amongst researchers on the best measure of privacy loss for consumers with AMI. The authors in \cite{Arzamasov2020} found that the perceived levels of SM privacy are greatly affected by the different privacy measurement methods, and that the privacy guarantees given by protection schemes based on these various methods are also highly dependent upon the underlying load profiles. Nevertheless, one widely accepted measure for privacy loss stemming from an adversary that has access to the individual grid-visible load profiles captured by the SMs is mutual information (MI) \cite{Yang2012, Giaconi2018a}. A method to tractably approximate MI in an optimisation problem for a single consumer  is proposed in \cite{Chin2017}. In this work, we adapt the MI approximate developed in \cite{Chin2017} as the privacy loss objective for multiple privacy-conscious consumers in the RDSA scheme. Let $X_l \in \mathcal{X}_l$ and $Y_l \in \mathcal{Y}_l$ denote the random variables modelling the instantaneous consumer load and the total grid load, respectively. For the purpose of estimating the MI only, we assume that these random variables are discrete and have finite support. In particular,  $\mathcal{\tilX}_l = \{x_l^0,x_l^1,\cdots,x_l^m\}$, $\mathcal{\tilY}_l = \{y_l^0,y_l^1,\cdots,y_l^n\}$, where $m$ and $n$ are the number of bins used to quantise the consumer and grid loads, respectively, and $I(X_l;Y_l)\approx I(\tilX_l;\tilY_l)$. The MI function for discrete random variables is given by
\begin{equation*}
    I(\tilX_l;\tilY_l) :=   \sum_{x_l \in \mathcal{\tilX}_l} \sum_{y_l \in \mathcal{\tilY}_l} p_{\tilX_l,\tilY_l}(x_l,y_l) \log \frac{p_{\tilX_l,\tilY_l}(x_l,y_l)}{p_{\tilX_l}(x_l)p_{\tilY_l}(y_l)},
\end{equation*} 
where $p_{\tilX_l,\tilY_l},~ p_{\tilX_l}$, and $p_{\tilY_l}$ are the approximated joint and marginal probability distribution functions (PDFs) of the random variables. In order to formulate MI as a function of the next grid load realisations, binary variables $\mathbf{z}_{l} = \{z_l^{ij} \}_{i = 1, j=1}^{m, n} \in \mathcal{Z}_l = \{0,1\}^{m \times n}$ are introduced in \cite{Chin2017} to relate the grid load variable, $y_l$ to its statistics. More specifically, the variables $z_l^{ij}$ count the frequency of $y_l$ when estimating its statistics using the histogram method. Given that the value of $x_l$ falls in the $i$-th bin, for each value of $y_l$, there exists exactly one non-zero $z_l^{ij}$ representing the bin where $y_l$ falls. Using $\mathbf{z}_{l}$ as the optimisation variables, an approximation of the MI function, $\tilI(\mathbf{z}_l)$ is then formulated as an optimisation objective\cite{Chin2017}. A brief summary on the derivation of $\tilI(\mathbf{z}_l)$ is provided in online Appendix A \cite{Chin2019}; further details can be found in \cite{Chin2017}. Here, we relax the binary constraints on $\mathbf{z}_{l}$, \emph{i.e.} $\mathbf{z}_{l} \in \mathcal{Z}'_l = [0,1]^{m \times n}$ and allow that for each value of $y_l$, there exists a set of non-zero $z_l^{ij}$ within the feasible set. Effectively, this relaxes the constraints imposed on $y_l$ by  $\mathbf{z}_{l}$, thereby affecting the statistics perceived by the controller, and hence, its privacy protection performance \cite{Chin2020}. Define then the set $\mathcal{F}'_l$ as the set of all $(y_l,\mathbf{z}_{l})$ that satisfies the constraints \eqref{eq:OriConStart} to \eqref{eq:OriConEnd}, in addition to the following constraints:

\begin{alignat}{2}
&\sum_{j=1}^{n} z^{i^*j}_l = 1 \\
& z^{ij}_l = 0 ~&&, \forall~ i \neq i^*\\
&\sum_{j=1}^{n} z^{i^*j}_l y^{j-1}_l \leq ~y_l < \sum_{j=1}^{n} z^{i^*j}_l y^{j}_l  ~ &&, \label{eq:ZtoY}
\end{alignat}
where $i^*$ is the index corresponding to the given value of $x_l$. Constraint \eqref{eq:ZtoY} links the grid load to its PDF estimate. Furthermore, let $\Phi(y_l) = \tilde{I}(\mathbf{z}_l)$ for any $\mathbf{z}_l$ satisfying $(y_l, \mathbf{z}_l) \in \mathcal{F}'_l$. $\tilde{I}(\mathbf{z}_l)$ is a quadratic form that is strongly convex for $m>1$. This can be implied from \cite{Chin2017} by relaxing the binary constraints and limiting the optimisation program to a single time step, and can be shown by analysing its algebraically manipulated form (see online Appendix B \cite{Chin2019}). Note that $\tilde{I}(\mathbf{z}_l)$ is only strongly convex for $m > 1$. Furthermore, the gradient of $\tilde{I}(\mathbf{z}_l)$ is bounded, thereby making $\tilde{I}(\mathbf{z}_l)$ Lipschitz continuous. These properties will be used later to apply the online gradient descent method to solve \eqref{eq:objective}.

\subsection{Ancillary Service Provision}
We consider the tracking of a target aggregate real power load profile as the ancillary service objective, and penalise deviations from this target profile, \emph{i.e.}, 
\begin{equation}\label{eq:gridservice}
    \Gamma(\textbf{y}) :=  \left \| \left(\sum_{l=1}^{N} y_l\right) - \bar{y} \right \|^2_2 ~,
\end{equation}
where $\bar{y}$ is the target profile, and $\mathbf{y} = \{y_1,y_2,\ldots,y_l \}$. This target profile can be shaped to provide services such as peak shaving, grid balancing, and congestion alleviation, or simply just to follow a planned consumption profile. Moreover, in well designed systems, power flow constraints arising from the provision of ancillary services by DERs are considered during their pre-qualification stage. These constraints are then taken into account when aggregating them for service provision; \emph{i.e.}, the aggregation groups can be defined such that participants within a group face similar constraints, which are then accounted for when sending out service requests. Note that one could also track an additional target reactive power profile in order to provide voltage support, but this is left as the subject of future work.

\section{Projected Online Gradient Descent} \label{solutionmethod}
The overall optimisation problem is now given by
\begin{equation}
\begin{split} \label{eq:optimisation}
& \underset{y_l,\mathbf{z}_l, s_l^+, s_l^-}{\mbox{minimise}} \quad \sum_{l=1}^{N} \left\{ \Lambda(y_{l}) + \mu_l \tilI(\mathbf{z}_l) \right\} + \rho \Gamma(\textbf{y})\\
& \mbox{subject to} \quad  (y_l, \mathbf{z}_l, s_l^+, s_l^-) \in \mathcal{F}'_l, \quad l = 1, \ldots, N,
\end{split}
\end{equation}
which can be solved optimally using an offline centralised controller by the aggregator if real-time high-resolution SM measurements and high-bandwidth communication links are available. However, given that current AMI deployments poll for readings at intermittent intervals due to communication infrastructure bandwidth constraints, this solution method remains impractical. Moreover, this would also require that private information from all consumers, \emph{i.e}, their actual consumer load, privacy preferences and day-ahead schedules, be revealed to the aggregator, invalidating the privacy protection objective. Alternatively, \eqref{eq:optimisation} can be solved using an offline distributed algorithm in the presence of two-way high-bandwidth communications infrastructure between scheme participants, which is again impractical for current grids. Therefore, we propose solving \eqref{eq:optimisation} using a distributed feedback-based online gradient descent solution method based on the general real-time feedback-based optimisation framework proposed in \cite{Dall'Anese2017}. The algorithm replaces the coupling of the grid load variables across consumers in \eqref{eq:gridservice} with the latest real-time aggregated consumption measurement $\hat{y}$ at the point of common coupling:
\begin{equation*}
    \sum_{l=1}^{N} y_l \approx  \hat{y}.
\end{equation*}

Hence, only the value $\| \hat{y} - \bar{y} \|^2_2$ is communicated in real-time, and is treated as a constant when computing the gradients. This makes \eqref{eq:optimisation} separable and solvable locally, overcoming the lack of real-time SM data and mitigating privacy concerns. The availability of $\hat{y}$ can readily be assumed given the increasing adoption of phasor measurement units (PMUs), and that transformers or busbars can easily be outfitted with a high-frequency measurement device. The use of $\hat{y}$ leads to a lagged and sub-optimal solution at each time step. However, by assuming that high-speed control actions can be actuated faster than the time-varying nature of \eqref{eq:optimisation}, the distributed algorithm is shown to converge to a centralised solution in \cite{Dall'Anese2017}, provided that the optimisation problem is strongly convex. 

Recall that \eqref{eq:optimisation} is strongly convex for $m>1$. Nevertheless, we add a regularisation term $\|\mathbf{h}\|^2_2$ with a small coefficient $\sigma_2/2$, where $\mathbf{h}$ is the vector of all the optimisation variables; this ensures that \eqref{eq:optimisation} is at least $\sigma_2$-strongly convex in all cases (e.g., when $m = 1$, or when extending the algorithm to multiple time steps). It is easy to see that $\|\mathbf{h}\|^2_2$, which is the sum of the squares of the variables, is separable. Accordingly, let\footnote{for simplicity in deriving the gradient} $\rho = \sigma_1/2$, and
\begin{equation*}
    f(y_l,\mathbf{z}_l)= \sum_{l=1}^{N} \left\{ \Lambda(y_{l}) + \mu_l \tilde{I}(\mathbf{z}_l) \right\} + \frac{\sigma_1}{2} \Gamma(\textbf{y}) + \frac{\sigma_2}{2}\|\mathbf{h}\|^2_2 ~.
\end{equation*}
We now solve the following quadratic program,
\begin{equation}
\begin{split} \label{eq:distopt}
& \underset{y_l,\mathbf{z}_l, s_l^+, s_l^-}{\mbox{minimise}} \quad f(y_l,\mathbf{z}_l)\\
& \mbox{subject to} \quad  (y_l, \mathbf{z}_l, s_l^+, s_l^-) \in \mathcal{F}_l' ~,  
\end{split}
\end{equation}
locally at each HEMS by first taking a gradient descent step:
\begin{alignat}{2}
    \tilde{s}^{+}_{l,t} &= &&~ s^{+}_{l,t-1} - r\nabla_{s^+_{l,t}} f(y_l,\mathbf{z}_l) \label{eq:s+int} \\
    \tilde{s}^{-}_{l,t} &= &&~ s^{-}_{l,t-1} - r\nabla_{s^-_{l,t}} f(y_l,\mathbf{z}_l) \label{eq:s-int} \\
    \tilde{z}^{i^*j}_{l,t} &=  &&~ z^{i^*j}_{l,t-1} - r\nabla_{z_{l,t}^{i^*j}} f(y_l,\mathbf{z}_l), ~j = 1,\ldots,n, \label{eq:zint}
\end{alignat}
where $r$ is the gradient descent step size, and $\nabla_A$ is the gradient with respect to $A$. The final solution is then obtained by projecting the interim solution onto the feasible set, \emph{i.e.}, 
\begin{equation}
 [{s}^{+}_{l,t},{s}^{-}_{l,t},{z}^{i^*j}_{l,t}]^\mathsf{T} = ~ \text{proj}_{\mathcal{F'}_l}[\tilde{s}^{+}_{l,t},\tilde{s}^{-}_{l,t},\tilde{z}^{i^*j}_{l,t}]^\mathsf{T} ~.   \label{eq:project}
\end{equation}
Only variables $z^{i^*j}_{l}$ are updated at each time step, with variables $z^{ij}_{l}$ for $i \neq i^*$ being treated as zero when computing the gradients for $z^{i^*j}_{l}$. Note that the actual values of $z^{ij}_{l}, ~i \neq i^*$ are not re-initialised as zero, and are kept for future time steps in order to ensure convergence. The gradients $\nabla_{s^+_{l,t}} f(y_l,\mathbf{z}_l)$, $\nabla_{s^-_{l,t}} f(y_l,\mathbf{z}_l)$, and $\nabla_{z_{l,t}^{i^*j}} f(y_l,\mathbf{z}_l)$ are derived from \eqref{eq:powerbalance} and \eqref{eq:distopt}, then computed by substituting for $\hat{y}$; see online Appendix B for the details on the gradients \cite{Chin2019}.


\subsection{Battery Modelling}
The convex modelling of realistic batteries while avoiding physically infeasible but optimal decisions due to simultaneous charging and discharging remains a research challenge. There are numerous modelling methods, e.g., using binary variables, quadratic constraints, or penalising battery use, but they are either non-convex, or are inapplicable because simultaneous charging and discharging is allowed and is optimal for \eqref{eq:optimisation} in some scenarios. To circumvent this issue, we assume that the battery is unable to go directly from charging to discharging, \emph{i.e}, it must go through ``zero",
\begin{equation*}
    (s^{+} > 0, s^{-}=0) \rightarrow (s^{+} = 0, s^{-}=0) \rightarrow (s^{+} = 0, s^{-} > 0)~,
\end{equation*}
and vice versa. This is a reasonable assumption given sufficiently fast control actions and limitations on certain power converter designs. 

The proposed distributed projected online gradient descent algorithm incorporating this battery modelling work-around is summarised in Algorithm \ref{alg:PGD}. We note that the convergence of this algorithm is guaranteed provided that the step size $r$ in \eqref{eq:s+int} - \eqref{eq:zint} is small enough, as the optimisation problem defined by \eqref{eq:distopt} satisfies the conditions identified in \cite{Dall'Anese2017}; see \cite{Dall'Anese2017} for details. The feasibility of the optimisation problem is also guaranteed as privacy requirements are formulated as a term in the objective function instead of a hard constraint.
\begin{algorithm}
\caption{algorithm for solving \eqref{eq:distopt} at time $t$}\label{alg:PGD}
\begin{algorithmic}[1]
\State obtain aggregated load measurement $\hat{y}_{t-1}$ 
\State obtain target load $\bar{y}_{t}$
\State compute $\| \hat{y}_{t-1} - \bar{y}_{t} \|^2_2$ and broadcast to consumers
\For {consumer 1 to N}
\State obtain load forecast $x_{l,t}$ and battery state $e_{l,t}$
\If {charge flag = true} 
\State compute $s^{+}_{l,t}$ and $z^{i^*j}_{l,t}$ using \eqref{eq:s+int}, \eqref{eq:zint} and \eqref{eq:project}
    \If {$s^{+}_{l,t} \geq 0$}
        \State actuate $s^{+}_{l,t}$ and update $z^{i^*j}_{l,t}$
    \Else
        \State charge flag = false
        \State actuate $s^{+}_{l,t} = 0$ and update $z^{i^*j}_{l,t}$
    \EndIf
\Else
\State compute $s^{-}_{l,t}$ and $z^{i^*j}_{l,t}$ using \eqref{eq:s-int}, \eqref{eq:zint} and \eqref{eq:project}
    \If {$s^{-}_{l,t} \geq 0$}
        \State actuate $s^{-}_{l,t}$ and update $z^{i^*j}_{l,t}$
    \Else
        \State charge flag = true
        \State actuate $s^{-}_{l,t} = 0$ and update $z^{i^*j}_{l,t}$
    \EndIf
\EndIf
\State update constants ($a^{ij}_l, ~b^{j}_l, ~c^{i}_l$) used in the MI 
\State approximation (see Appendix A in \cite{Chin2019} for details)
\EndFor
\State \textbf{advance} to $t+1$
\end{algorithmic}
\end{algorithm}
\section{Numerical Experiments}
\label{imp}
The proposed scheme is tested using 1 Hz smart meter data taken from the ECO dataset \cite{Kleiminger2015}. As there are only six houses in this dataset, we emulated more consumers in the RDSA scheme by drawing data from multiple days over the period between $26$ August and $9$ September, 2012 from the six households. This results in a dataset with homogeneous load profiles, which may decrease target aggregate load tracking performance. Given more heterogeneous load profiles (as in reality), the resultant aggregate load would likely be smoother as load peaks are compensated by load troughs, allowing for better tracking.
 
\subsection{Simulation Setup}
Each household in the RDSA scheme was assigned a random price of privacy loss, $1 \leq \mu_l \leq 9$, to mimic the behaviour of multiple real households. To enforce the assumption that control actions are actuated faster than the time varying nature of \eqref{eq:distopt}, we assume that the local high frequency SM measurements have a resolution of 0.2 Hz, and that the target grid signal also varies every 5 seconds, while the aggregated load measurements are available every second. These are realistic assumptions, as local HEMS controllers obtain measurements from smart meters deployed locally at each household, which have the ability to sample at 2kHz or higher; and aggregators receiving continuous real-time reserve provision signals may choose to broadcast a more stable signal, so long as they meet the requirements in their service contracts with the grid operator. 

Each household tracks an arbitrary day-ahead schedule, $y^\textit{ref}_l$, computed using a single multi-time step optimisation for energy costs and MI privacy, at half-hourly resolution with the MDPC algorithm described in \cite{Chin2017}, and on a two-tier time-of-use energy tariff. Here, we define $\hat{y} = \sum_{l=1}^{N} y_l$ for simplicity, omitting grid losses, and assume that grid in-feed is not allowed, \emph{i.e}, $y_l^{min} = 0$. Depending on network topology and size, the grid losses may impact the RDSA scheme's performance in reality, but this is outside the scope of this paper. The general system setup at each consumer household is summarised in Table \ref{tab:consumersystem}, where battery parameters were taken from the Tesla Powerwall v1. 

The target aggregate load $\bar{y}$ is generated using the aggregated day-ahead consumer schedules as a base, considering total RDSA reserve capacity $\bar{\gamma} = N \gamma$. For ease of assessment, this is designed such that the target load is energy neutral with respect to the aggregate day-ahead schedule within each half-hourly interval. When generating such a reference curve, we took into account battery losses (max bias of 0.1\% of reserve capacity), and ensured that the aggregate reserves are sufficient to meet the ancillary service requests approximately 99.7\% of the time.

Unless otherwise stated, the simulation parameters are as listed in Table \ref{tab:simparam} where applicable. For ease of comparison and simplicity, we set aside battery energy capacity corresponding to $\gamma$ at each consumer household in their day-ahead schedules, and assume that the battery model in the optimisation problem is accurate. 
\begin{table}
\renewcommand{\arraystretch}{1.1}
\centering
\caption{Individual Household System Parameters}
\label{tab:consumersystem}
\begin{tabular}{|l|c|} \hline 
Real-Time PDF Estimation Sample Size, $K+1$   & $901$ \\  
Number of $\mathcal{X}$ Bins, $m$           & $15$ \\    
Number of $\mathcal{Y}$ Bins, $n$           & $15$ \\     
Additive Smoothing, $\varepsilon$           & $0.1$\\    
Battery Capacity                            & $6.4$ kWh\\ 
Battery Power                               & $3.3$ kW\\
One-way Battery Efficiency, $\eta$          & $96$ \%\\
RDSA Reserve Capacity, $\gamma$             & $0.15$ kW\\
\hline 
\end{tabular}
\end{table}
\begin{table}
\renewcommand{\arraystretch}{1.1}
\centering
\caption{General Simulation Parameters}
\label{tab:simparam}
\begin{tabular}{|l|c|} \hline 
Initial Battery State of Charge             & $3.2$ kWh\\
Ancillary Service Coefficient, $\sigma_1$   & $5$ \\
Regularisation Coefficient, $\sigma_2$      & $1e^{-4}$ \\
Descent Step Size, $r$                      & $0.012$ \\
No. of Consumers, $N$                       & $20$ \\
\hline 
\end{tabular}
\end{table}
\begin{table}
\renewcommand{\arraystretch}{1.3}
\centering
\caption{Variables used in the optimisation problem at time $t$}
\label{tab:availableinfo}
\begin{tabular}{|l|c|c|} \hline 
                        &Distributed        &Centralised \\ \hline
Load Forecast           &$x_{l,t-1}$        &$x_{l,t}$  \\
Target Aggregate Load   &$\bar{y}_{l,t}$    &$\bar{y}_{l,t}$ \\
Day-Ahead Schedule      &$y^\textit{ref}_{l,t}$    &$y^\textit{ref}_{l,t}$ \\
Aggregate Load          &$\hat{y}_{l,t-1}$  &$\sum_{l=1}^{N}y_{l,t}$ \\
\hline 
\end{tabular}
\end{table}

The proposed distributed algorithm (abbreviated as Dist.POGD in this section) is compared against an ideal centralised solution for \eqref{eq:distopt}, both with and without relaxing the binary constraints on $\mathbf{z}_l$ by modelling it in YALMIP \cite{Lofberg2004}, and solving it with the Gurobi solver. For the distributed algorithm, we assume a persistent (rather than perfect) forecast for $x_l$, \emph{i.e.}, use the latest high-frequency SM measurement for computing the gradient step. For the centralised solution, a perfect forecast was used instead in order to provide a better benchmark. Note that $\sum_{l=1}^{N} y_l$ is not substituted with $\hat{y}$ in the centralised solution. Table \ref{tab:availableinfo} summarises the information used in the respective solution methods at time $t$.

\subsection{Visualisation of Aggregate Loads}
\begin{figure*}
    \centering
    \subfloat[Target grid load and aggregated grid loads]{
    \includegraphics[trim=1.9cm 9.5cm 1.27cm 10.5cm, clip=true, width=1.88\columnwidth]{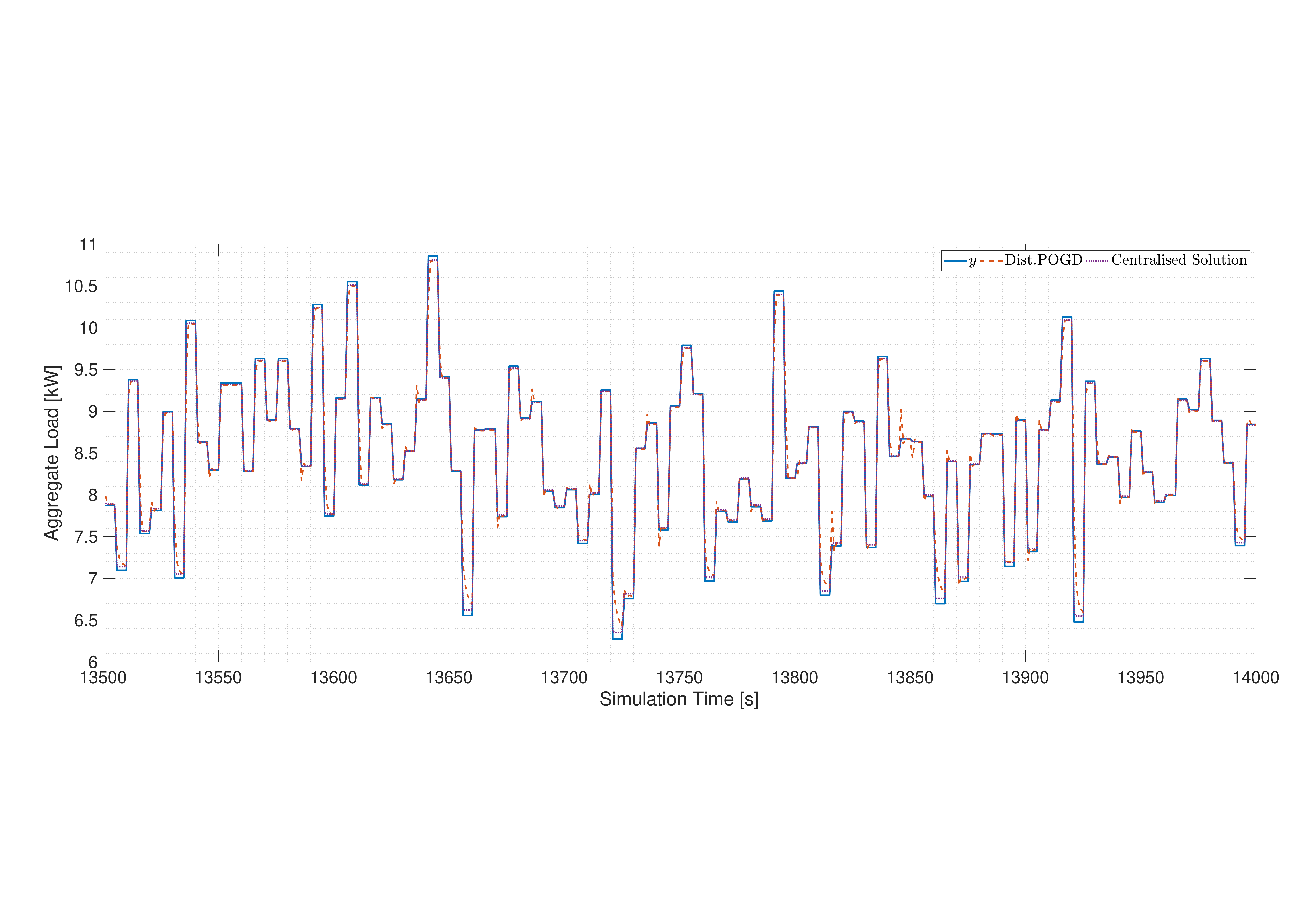} \vspace{-0.0cm}
    \label{fig:RefLoadProfiles}
    }\\ \vspace{-0.2cm}
    \subfloat[Aggregated consumer load]{
    \includegraphics[trim=2cm 11cm 1.30cm 21cm, clip=true, width=1.88\columnwidth]{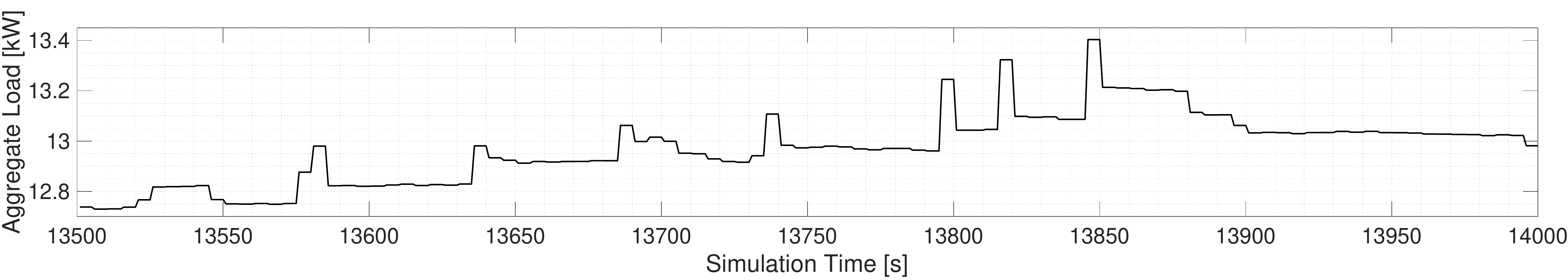}
    \label{fig:AggregateConsumerLoad}
    }
    \caption{Target grid load, and aggregated grid and consumer loads}
    \label{fig:ReferenceLoadProfiles}
\end{figure*}

Fig. \ref{fig:ReferenceLoadProfiles} illustrates the target aggregate load profile, the aggregated grid loads from Dist.POGD and a centralised solution with relaxed binary constraints, and the total consumer load. Within the specific illustrated period, the batteries are discharging, resulting in grid loads less than the total consumer load; but the converse can be true in other periods. Using the parameters in Table \ref{tab:simparam} and the selected energy and privacy loss prices, the Dist.POGD solution converges close to the ideal centralised solution, as seen in Fig. \ref{fig:ReferenceLoadProfiles}, and overshoots only when there is a significant consumer load change. The performance of Dist.POGD is affected by the choice of parameters in Table \ref{tab:simparam}. These parameters have to be chosen empirically in a real system, as consumer privacy preferences would be unknown to the aggregator.    


\subsection{Evaluation Metrics}
Numerically, we evaluate the performance of the algorithms over a period of three and a half hours (number of samples $K_s = 12600$; simulate four hours, discard the first half-hour for initialisation purposes) with the following metrics. For ancillary service provision, the algorithms are evaluated based on the normalised root mean square error (NRMSE) between $\hat{y}$ and $\bar{y}$, given by 
\begin{equation*}
\text{NRMSE} := \frac{\sqrt{\frac{1}{K_{s}}\sum_{k=1}^{K_{s}}(\hat{y}_k-\bar{y}_k)^2}}{\bar{y}^{mean}} \times 100\% ~,
\end{equation*}
\begin{equation*}
\bar{y}^{mean} = \frac{1}{K_{s}} \sum_{k=1}^{K_{s}}  \bar{y}_k ~,  
\end{equation*}
and the mean absolute percentage error (MAPE),
\begin{equation*}
\text{MAPE} := \frac{100\%}{K_{s}} \sum_{k=1}^{K_{s}} \left| \frac{\hat{y}_k - \bar{y}_k}{\bar{y}_k} \right| ~.
\end{equation*} 
MAPE gives the average of the errors at each time $t$, while NRMSE emphasises large deviations from the target load, which are undesirable, and reflects the target load tracking error relative to the overall mean. The consumer preference (day-ahead schedule tracking) is evaluated by computing its normalised mean absolute error (NMAE), \emph{i.e.}, the absolute error as a percentage of the maximum grid load, 
\begin{equation*}
\text{NMAE} := \frac{100\%}{K_{s} y_l^{max}} \sum_{k=1}^{K_{s}} \left| y_{l,k} - y^{ref}_{l,k} \right| ~.
\end{equation*}
We used an approximate MI function in the objective function of the optimisation problem. However, for evaluation purposes we estimate the average MI, assuming identical and independently distributed random variables (denoted as $I_{iid}$), and stationary first-order Markov processes (denoted as $I_\textit{mk}$), as described in \cite{Tan2017,Chin2018}. Such evaluation requires the computation of $p_{X,Y}(x,y) \log [p_{X,Y}(x,y)/p_X(x)p_Y(y)]$, which in the case of $p_{X,Y}(x,y)$ being zero, is set to zero to avoid computing $\log 0$. The $I_\textit{mk}$ captures some of the time correlation between the consumer and grid loads, which is neglected when calculating the $I_{iid}$ \cite{Chin2018}. Note that while modelling the $X_l$ and $Y_l$ as time-varying Markov processes would yield more accurate MI estimates, there are insufficient samples from this simulation for its application. 

The information that is contained within a consumer's load profile has been shown to be dependent on the measurement frequency \cite{Rousseau2020}. Thus, the privacy risks stemming from high-frequency SM measurements are also time resolution dependent, \emph{i.e}, the resolution at which consumers are metered influences their privacy risks \cite{Eibl2015}. Given that 1-minute metering resolution is more likely to be deployed than the 1-second time resolution of the controller, the MI is also assessed at this lower time resolution to better understand the realistic SM privacy loss. 

While there is a lack of literature that studies SM privacy protection considering access to side information, the presence of such information should not be ignored due to its potential impact on the performance of privacy protection schemes. In a recent preliminary work, even the presence of mundane side information (day of the week) has been shown to degrade the performance of two different learning-based SMDM privacy protection mechanisms \cite{Shateri2020}. While adversarial attack models are not the focus of this paper, for the sake of completeness, we also considered privacy loss in the presence of some simple side information, which may increase the information leakage to the adversary. In our case study, one could reasonably assume that an adversary (possibly the aggregator) may have the following side information: the ancillary service provision signal $\| \hat{y}_{t-1} - \bar{y}_{t} \|^2_2$, and each consumer's day-ahead schedule $y_l^\textit{ref}$. Each consumer's grid load on the other hand is composed of their day-ahead schedule, deviations due to privacy-protection, ancillary service request, and consumer load deviations from day-ahead forecasts. Hence, a naive, but simple assessment of such risk is conducted by estimating the MI of the grid load profiles with the ancillary service portion removed (denoted as \textit{GS} or $y_l^\text{G}$), with the day-ahead schedule removed (dentoed as \textit{DA} or $y_l^\text{D}$), and with both subtracted (denoted as \textit{DAGS} or $y_l^\text{DG}$): 
\begin{alignat*}{1}
y_l^\text{G} &= \, y_l - \overbrace{(\hat{y} - \bar{y})\frac{y_{l}^\text{mean}}{\sum_{l=1}^{N}y_{l}^\text{mean}}}^{\text{ancillary service request}} ~,\\
y_{l}^\text{mean} & = \, \frac{1}{k_s} \sum_{k = 1}^{k_s} y_{l,k} ~, \\
y_l^\text{D} & = \, y_l - y_l^\text{ref} ~\\
y_l^\text{DG} & = \, y_l^\text{G}- y_l^\text{ref} ~.
\end{alignat*}

\begin{figure} 
    \centering
    \includegraphics[trim=4cm 2.5cm 4cm 3.5cm, clip=true, width=0.98\columnwidth]{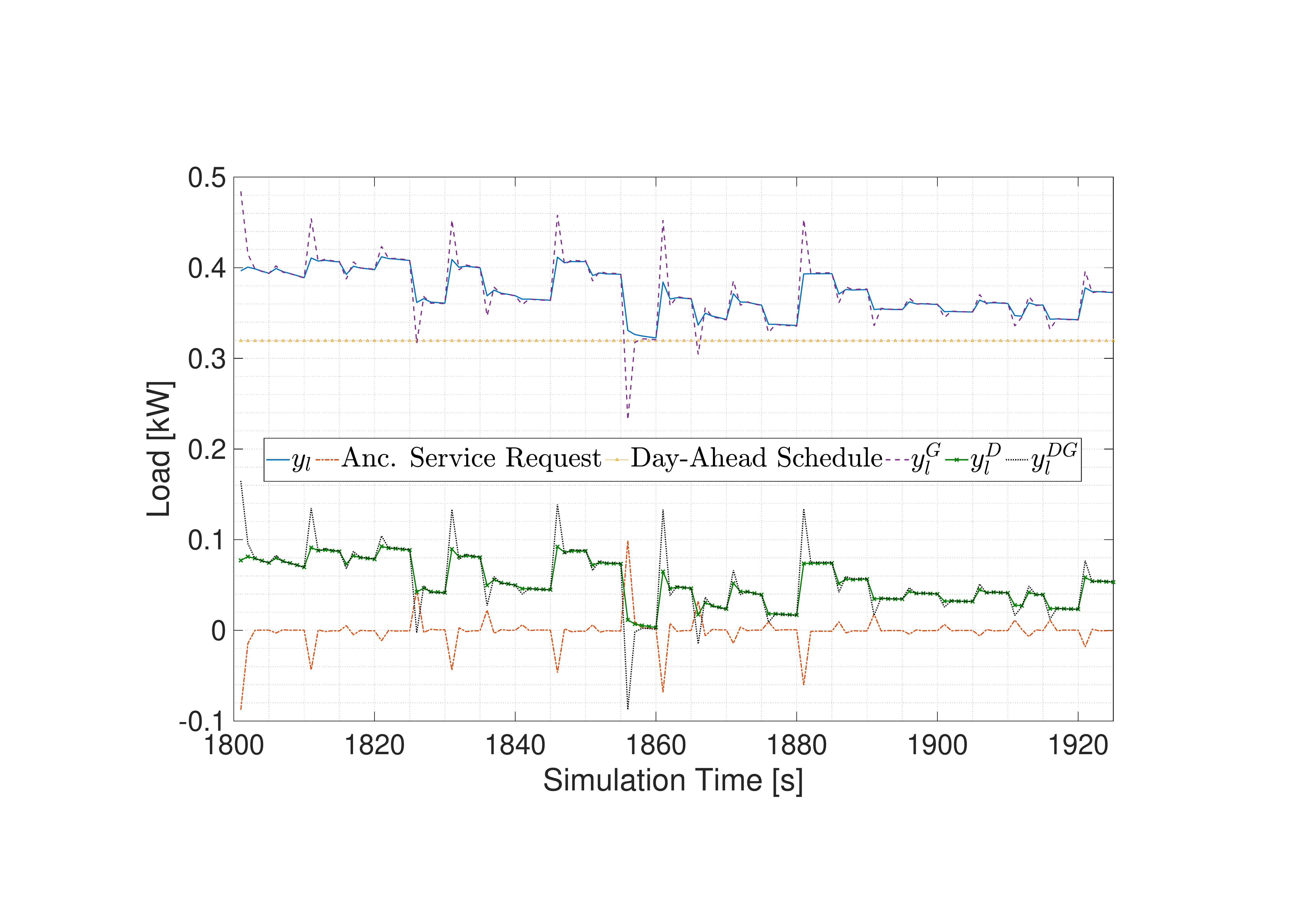} \vspace{-0.2cm}
    \caption{Grid load for consumer 1}
    \label{fig:CorrectGridLoad}
\end{figure}

Fig. \ref{fig:CorrectGridLoad} illustrates the original and adjusted grid loads for consumer $1$. Note that the ancillary service request does not equal the ancillary service provision, and that $y_l^\text{DG}$ is the adversary's guess that does not reflect the actual privacy sensitive consumer load. The number of bins, $m$ and $n$, are kept the same as in the optimisation problem when computing the MI between $x_l$ and $y_l$. However, $n$ is adjusted when computing the MI for $y^\text{G}_l$,  $y^\text{D}_l$ and $y^\text{DG}_l$ to account for the increased range space; in order to keep a similar quantisation resolution. The increase in range space for $y^\text{G}_l$ and $y^\text{DG}_l$ can be seen in Fig. \ref{fig:CorrectGridLoad}; while the range space increase for $y^\text{D}_l$ occurs whenever there is a change in the day-ahead schedule (e.g., every half hourly in our case study). All else being equal, increasing the quantisation resolution generally leads to an increase in the MI estimate as discussed in \cite{Chin2017}.

\subsection{Results and Discussion}
Tables \ref{tab:TrackingUtility} and \ref{tab:MI} summarise the performance of Dist.POGD with different parameters, and the two centralised solutions. While the Dist.POGD solution converges close to the ideal centralised solution as seen in Fig. \ref{fig:ReferenceLoadProfiles}, there is a deterioration in the overall performance due to the time required for convergence, forecast error, and the replacement of $\hat{y}_{l,t-1}$ for $\sum_{l=1}^{N}y_{l,t}$. For the simulations, a persistent forecast was used in Dist.POGD instead of a perfect forecast. This allows for a better analysis of its performance if deployed in reality. Moreover, relaxing the binary constraints on $\mathbf{z}_l$ marginally improves the NRMSE, MAPE, average NMAE, and most of the MI estimates, showing that this relaxation does not significantly impact the performance of the MI approximate in the objective function for a single control action. Note, however, that this may not be true in general; in fact, the opposite might occur as shown in \cite{Chin2020}. 

One can also see an increase in information leakage if MI is assessed at a resolution lower than the controller's, e.g., at one minute instead of one second. However, this increase might not pertain to information that is masked by the controller at higher resolutions, e.g., information on the type of device (contained in low time resolution data) versus a device's condition (inferable from high frequency data). By definition mutual information measures the general interdependence between two statistical distributions. As such, it is immune to post-processing for the information it is designed to hide. However, this does not mean that the grid load does not carry different (unprotected) information about the consumer load that is inferable from lower time resolution (aggregated) data . Recall that time aggregation has significant impact on SM load profile characteristics and by extension, also on the information contained within them \cite{Rousseau2020}.

To gain a better understand the effect of time resolution (aggregation) on the privacy levels attained by privacy mechanism employed in this paper, Houses 1 and 11 were simulated and evaluated at different time resolutions, with the controller objective set to only protect privacy. Tables \ref{tab:H1MIOnly} and \ref{tab:H11MIOnly} show the results of the different MI estimates. Assuming that the battery controller is able to maintain a fixed value across higher time resolutions (\emph{i.e.}, hold a fixed grid load over the control action interval), it can be seen that evaluating privacy loss at time resolutions higher than those used for the control actions always led to less privacy-loss. This is intuitive as it is akin to load-levelling across the evaluation time interval. However, it is inconclusive to state that aggregating high time resolution measurements always leads to more privacy-loss; e.g., there is less IID MI when evaluating 1-minute simulations for House 1 at 5-minute resolutions. We believe that the effects on time aggregation during evaluation may be dependent on the amount of time-correlated information present in the underlying consumer load, but the investigation of this is left for future work.
\begin{table}
    \renewcommand{\arraystretch}{1.1}
    \centering
    \caption{House 1 Privacy Protection Only}
    \label{tab:H1MIOnly}
    \begin{tabular}{|l|c|c|c|c|c|c|} \hline 
        \multirow{2}{*}{}           & \multicolumn{6}{c|}{Evaluation Resolution}\\ 
        \cline{2-7} 
                  & \multicolumn{3}{c|}{IID MI}       & \multicolumn{3}{c|}{Markov MI} \\ 
           \hline   
           Sim. Reso.         & $1$ Sec     & $1$ Min     & $5$ Min     & $1$ Sec     & $1$ min     & $5$ Min \\ 
           \hline    
           $1$ Second              & \cellcolor{yellow} $0.068$    &  $0.067$         &  $0.092$         & \cellcolor{yellow} $0.011$         & $0.102$          & $0.393$\\
           \hline
           $1$ Minute              & $0.412$     & \cellcolor{yellow} $0.437$         & $0.358$          & $0.0030$          & \cellcolor{yellow} $0.220$          & $0.522$\\
           \hline
           $5$ Minutes             & $0.074$     & $0.079$          & \cellcolor{yellow} $0.105$         & $6.95$e$^\text{-4}$          & $0.038$           & \cellcolor{yellow} $0.195$\\ 
           \hline 
    \end{tabular}

    \renewcommand{\arraystretch}{1.1}
    \centering
    \caption{House 11 Privacy Protection Only}
    \label{tab:H11MIOnly}
    \begin{tabular}{|l|c|c|c|c|c|c|} \hline 
        \multirow{2}{*}{}           & \multicolumn{6}{c|}{Evaluation Resolution}\\ 
        \cline{2-7} 
               & \multicolumn{3}{c|}{IID MI}       & \multicolumn{3}{c|}{Markov MI} \\ 
        \hline   
        Sim. Reso.         & $1$ Sec     & $1$ Min     & $5$ Min     & $1$ Sec     & $1$ min     & $5$ Min \\ 
        \hline    
        $1$ Second              & \cellcolor{yellow} $0.681$    & $0.689$           & $0.736$          & \cellcolor{yellow} $0.031$          & $0.232$          & $0.309$\\
        \hline
        $1$ Minute              & $0.446$     & \cellcolor{yellow} $0.579$          & $0.571$          & $0.006$          & \cellcolor{yellow} $0.216$         & $0.320$\\
        \hline
        $5$ Minutes             & $0.361$     & $0.414$          & \cellcolor{yellow} $0.552$          & $0.004$         & $0.099$          & \cellcolor{yellow} $0.318$\\ 
        \hline 
    \end{tabular}
\end{table}

Note also that the 1-sec $I_\textit{mk}$ values are very small because the measure is unable to fully capture the time-correlated privacy leakage that it was designed for. The time-correlation of the load profiles is at least 5 seconds (5 sample intervals), which cannot be captured by modelling the loads as first-order Markov processes, a drawback that was previously highlighted in \cite{Chin2018}. 
    
Table \ref{tab:MIComp} summarises the MI estimates for the compensated grid loads using the reference Dist.POGD setup. In analysing the effects of the simple grid load compensation using the possible side information, we consider only added information leakage as this illustrates the loss of privacy in the presence of such information. Comparing Tables \ref{tab:MI} and \ref{tab:MIComp}, one can see that correcting for ancillary service request results in higher 1-sec $I_\textit{mk}$, and 1-min $I_\textit{mk}$ estimates, thus resulting in more leakage of time-correlated private information. While there is a slight increase in 1-min $I_{iid}$, the results are inconclusive. Having the day-ahead consumer schedule does not appear to increase information leakage in our experiments. This can be seen as there is no marked increase in information leakage when compensating for the day-ahead schedule alone (DA), or further compensating for it after removing the grid service request (DAGS). This shows that the ancillary service request signal constitutes sensitive side-information, while additional information on the day-ahead schedule may not exacerbate consumer privacy loss. Note that the day-ahead schedules in this case study are optimised for half-hourly privacy protection in addition to energy cost, which may explain the lack of sensitive information derivable from them. Further study on ways of incorporating the day-ahead schedules are needed for a conclusive verdict on their privacy sensitivity.

\begin{table}
\renewcommand{\arraystretch}{1.1}
\caption{Performance of Different Solution Methods and Parameters}
\label{tab:TrackingUtility}
\centering
\begin{tabular}{|c|c|c|c|}
\hline
                        & NRMSE         & MAPE          & Avg. NMAE \\
\hline 
Reference Dist.POGD          & $1.694\%$      & $0.7878\%$   & $4.699\%$   \\
\hline 
Centralised, binary     & $0.336\%$      & $0.295\%$    & $3.71\%$   \\
Centralised, relaxed    & $0.322\%$      & $0.283\%$    & $3.53\%$   \\
\hline 
15 Consumers            & $2.69\%$      & $1.61\%$      & $5.42\%$   \\
25 Consumers            & $1.51\%$      & $0.827\%$     & $4.32\%$   \\
\hline 
$r=0.008$               & $2.80\%$      & $1.63\%$      & $5.13\%$   \\
$r=0.016$               & $1.56\%$      & $0.914\%$     & $4.41\%$   \\
\hline
$\sigma_2=0$            & $1.691\%$      & $0.7873\%$     & $4.695\%$   \\
$\sigma_2=0.001$        & $1.694\%$      & $0.7883\%$     & $4.703\%$   \\
\hline
$\mu_l=0$               & $1.68\%$      & $0.777\%$     & $4.61\%$   \\
$11\leq\mu_l\leq19$     & $1.78\%$      & $0.866\%$     & $5.21\%$   \\
\hline
$K=600$                 & $1.70\%$      & $0.796\%$     & $4.71\%$   \\
$K=1200$                & $1.69\%$      & $0.785\%$     & $4.69\%$   \\
\hline
$\sigma_1=3$            & $3.17\%$      & $2.05\%$     & $4.41\%$   \\
$\sigma_1=7$            & $1.65\%$      & $0.978\%$     & $4.86\%$   \\
\hline
\end{tabular} \vspace{-0.3cm}
\end{table}

\begin{table}
	\renewcommand{\arraystretch}{1.1}
	\caption{MI Estimates for Different Solution Methods and Settings}
	\label{tab:MI}
	\vspace{0.1cm}
	\centering
	\begin{tabular}{|c|cc|cc|}
		\hline	
											&\multicolumn{2}{c|}{Avg. $1$ sec} 	&\multicolumn{2}{c|}{Avg. $1$ min}		\\
		\cline{2-5} 
											& $I_{iid}$       	& $I_\textit{mk}$         & $I_{iid}$       	& $I_\textit{mk}$         \\
		\hline
		Reference Dist.POGD          		& $0.248$   	& $0.020$           & $0.280$   	& $0.193$           \\
		\hline 
		Centralised, binary     			& $0.236$   	& $0.018$           & $0.277$   	& $0.140$           \\
		Centralised, relaxed    			& $0.221$   	& $0.015$           & $0.270$   	& $0.147$           \\
		\hline 
		$15$ Consumers            			& $0.301$   	& $0.018$           & $0.354$   	& $0.228$           \\
		$25$ Consumers            			& $0.223$   	& $0.020$           & $0.244$   	& $0.164$           \\
		\hline 
		$r_g=0.008$               			& $0.256$   	& $0.019$           & $0.300$   	& $0.194$           \\
		$r_g=0.016$               			& $0.247$   	& $0.022$           & $0.279$   	& $0.184$           \\
		\hline
		$\sigma_{\text{agg},2}=0$           & $0.248$   	& $0.020$           & $0.281$   	& $0.192$           \\
		$\sigma_{\text{agg},2}=0.001$       & $0.247$   	& $0.020$           & $0.281$   	& $0.192$           \\
		\hline
		$\mu_l=0$               			& $0.254$   	& $0.020$           & $0.287$   	& $0.194$           \\
		$11\leq\mu_l\leq19$     			& $0.237$   	& $0.020$           & $0.269$   	& $0.196$           \\
		\hline
		$K=600$                 			& $0.247$   	& $0.020$           & $0.276$   	& $0.196$           \\
		$K=1200$                			& $0.248$   	& $0.020$           & $0.282$   	& $0.196$           \\
		\hline
		$\sigma_{\text{agg},1}=3$           & $0.253$   	& $0.018$           & $0.294$   	& $0.189$           \\
		$\sigma_{\text{agg},1}=7$           & $0.250$   	& $0.023$           & $0.285$   	& $0.204$           \\
		\hline
	\end{tabular}
\end{table}

\begin{table}
	\renewcommand{\arraystretch}{1.1}
	\caption{MI Estimates for the Compensated Grid Load}
	\label{tab:MIComp}
	\vspace{0.1cm}
	\centering
	\begin{tabular}{|c|cc|cc|}
		\hline
		\multicolumn{5}{|c|}{\textbf{Reference Dist.POGD}}	\\
		\hline	
		~						&\multicolumn{2}{c|}{Avg. $1$ sec} 	&\multicolumn{2}{c|}{Avg. $1$ min}		\\
		\cline{2-5} 
		~						& $I_{iid}$       	& $I_\textit{mk}$         & $I_{iid}$       	& $I_\textit{mk}$     \\
		\hline
		GS						& $0.236$		& $0.039$			& $0.306$		& $0.249$		\\
		\hline
		DA						& $0.102$		& $0.022$			& $0.139$		& $0.224$		\\
		\hline
		DAGS					& $0.086$		& $0.033$			& $0.138$		& $0.225$		\\
		\hline
	\end{tabular}
\end{table}

\begin{figure}
    \centering
    \subfloat[15 consumers]{
    \includegraphics[trim=1.6cm 5.6cm 1.65cm 6.5cm, clip=true, width=0.98\columnwidth]{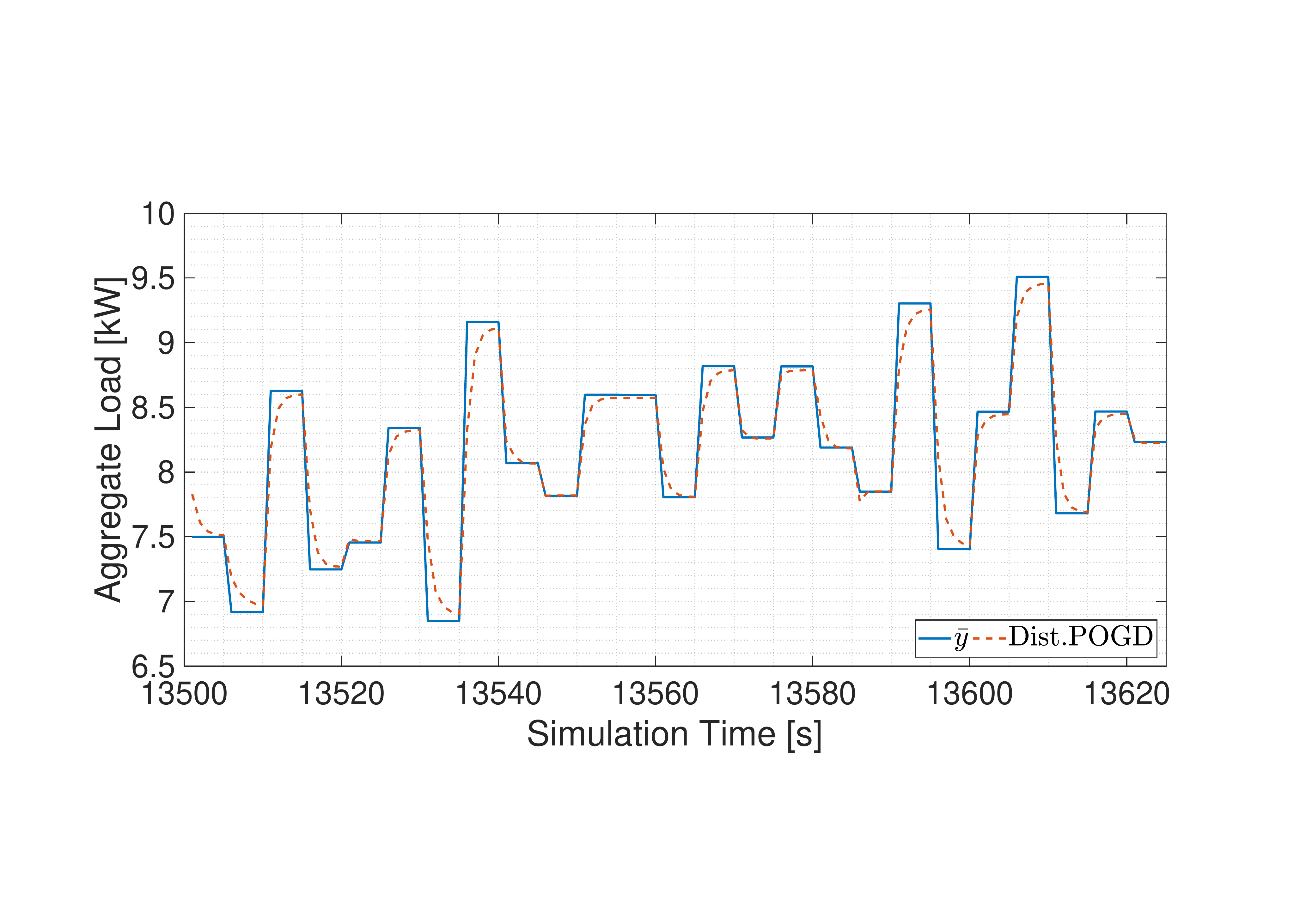}
    \label{fig:15Consumers}
    }\\ \vspace{-0.1cm}
    \subfloat[25 consumers]{
    \includegraphics[trim=1.6cm 5.6cm 1.5cm 6.5cm, clip=true, width=0.98\columnwidth]{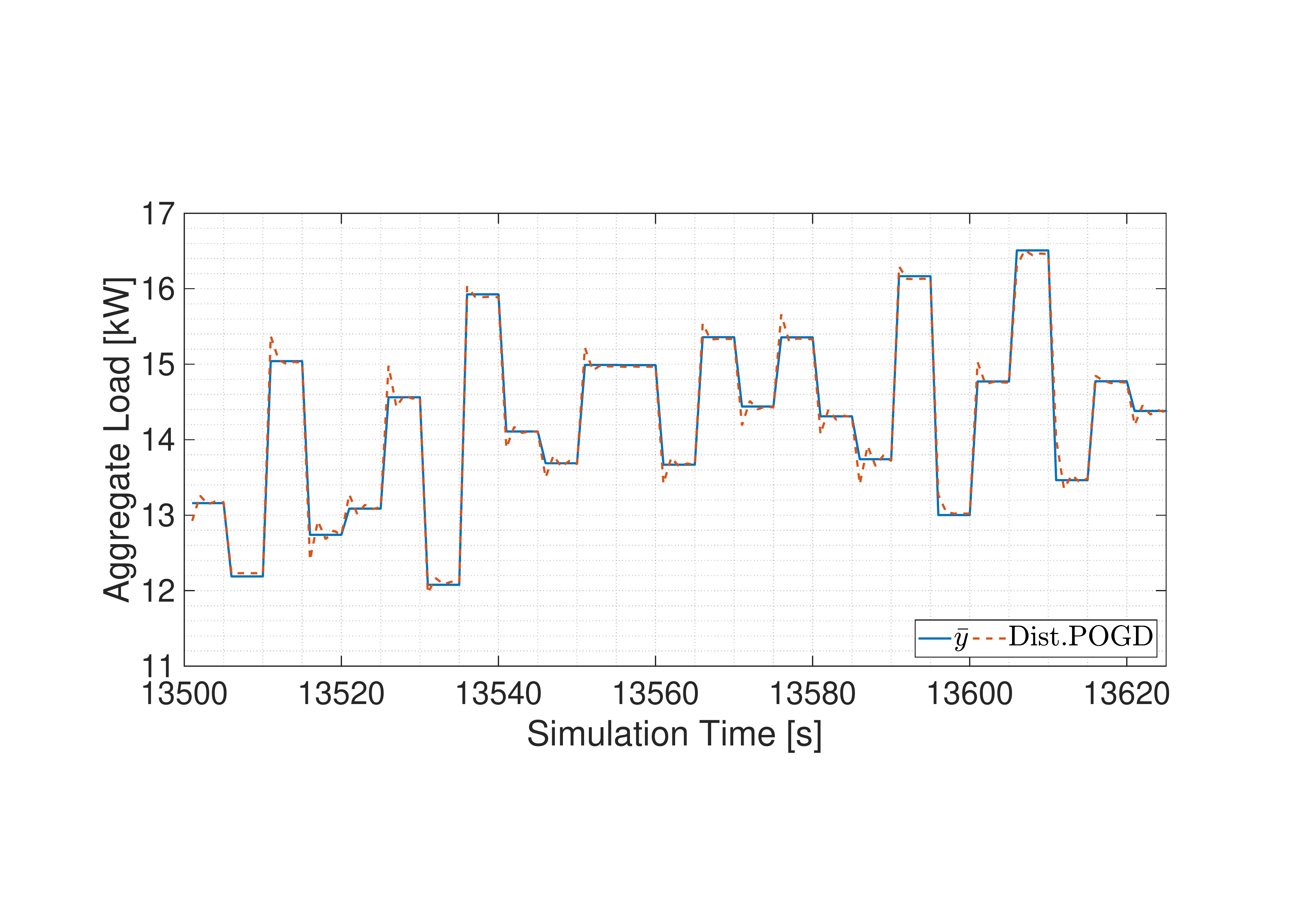}
    \label{fig:25Consumers}
    }
    \caption{Tracking performance with different aggregation sizes}
    \vspace{-0.1cm}\label{fig:ConsumerQuant}
\end{figure}

\begin{figure} 
    \centering
    \includegraphics[trim=1.6cm 4.7cm 1.6cm 6.5cm, clip=true, width=0.98\columnwidth]{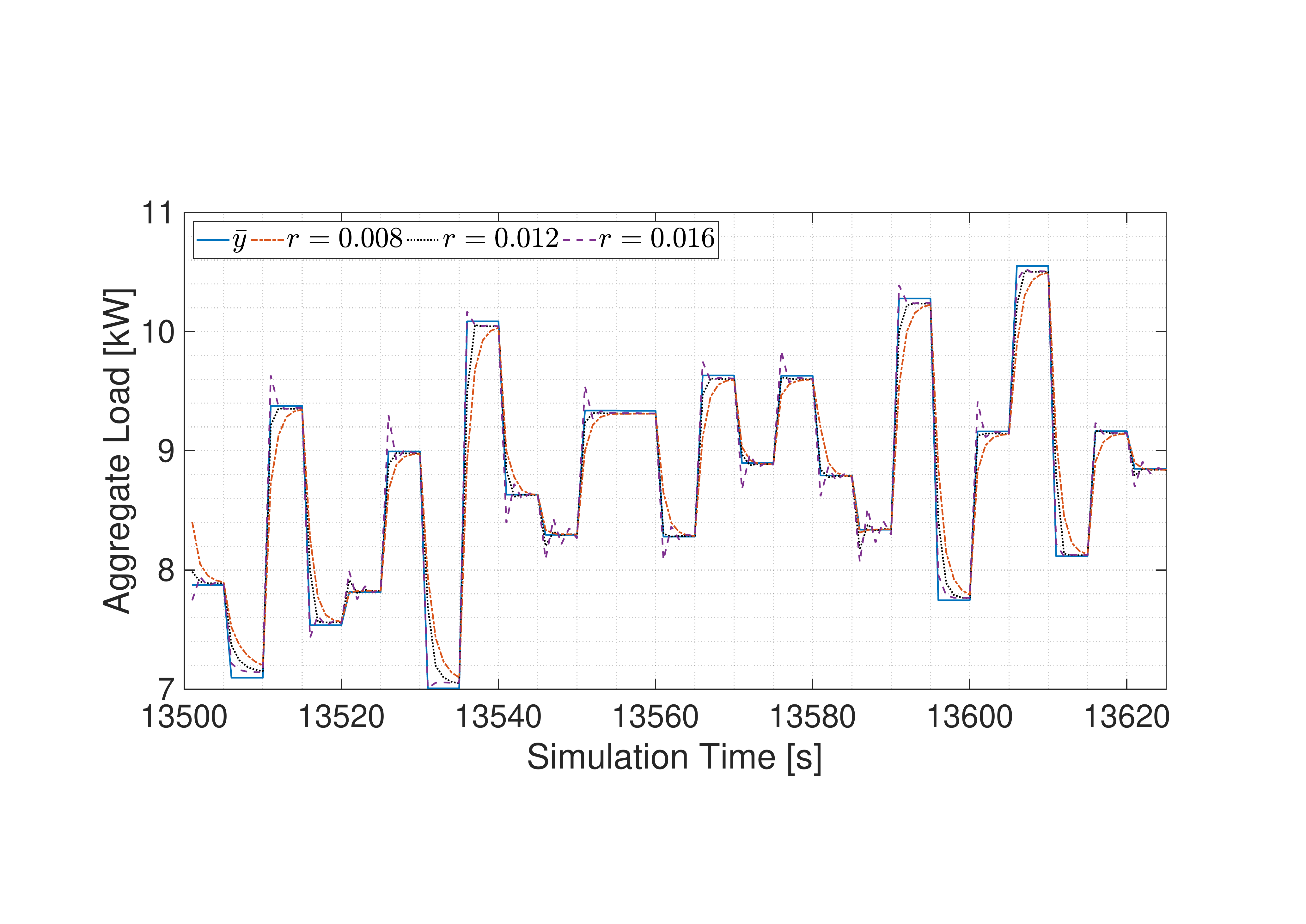} \vspace{-0.1cm}
    \caption{Comparison between different step sizes, $r$}
    \label{fig:StepSize}
    \vspace{-0.3cm}
\end{figure}

As discussed, the RDSA parameters are setup-dependent, and changing the number of consumers in the RDSA  while keeping all else constant affects the convergence rates and performance. This is illustrated for 25 and 15 consumers in Fig. \ref{fig:ConsumerQuant}. The choice of gradient descent step size $r$ affects the convergence rate of Dist.POGD; too high a value for $r$ leads to overshoots and instability, while too low a value hinders convergence to the optimal solution before the problem changes, as seen in Fig. \ref{fig:StepSize}. 1-Sec $I_\textit{mk}$ increases with larger step sizes (see Table \ref{tab:MI}), showing an increase in the leakage of time-correlated private information with larger $r$ values. On the other hand, the overall tracking performance improves and $I_{iid}$ decreases as the rate of convergence (and overshoots) increases (see Table \ref{tab:TrackingUtility} and Fig. \ref{fig:StepSize}). An ideal value for the coefficient $\sigma_2$ for the regularisation term $\|\mathbf{h}\|^2_2$ would result in minimal impact on the performance of the algorithm. As seen in Tables \ref{tab:TrackingUtility} and \ref{tab:MI}, Dist.POGD with $\sigma_2 = 1e^{-4}$ has similar performance to an algorithm without regularisation.

\begin{figure}
    \centering
    \includegraphics[trim=1.6cm 5.7cm 1.6cm 6.5cm, clip=true, width=0.98\columnwidth]{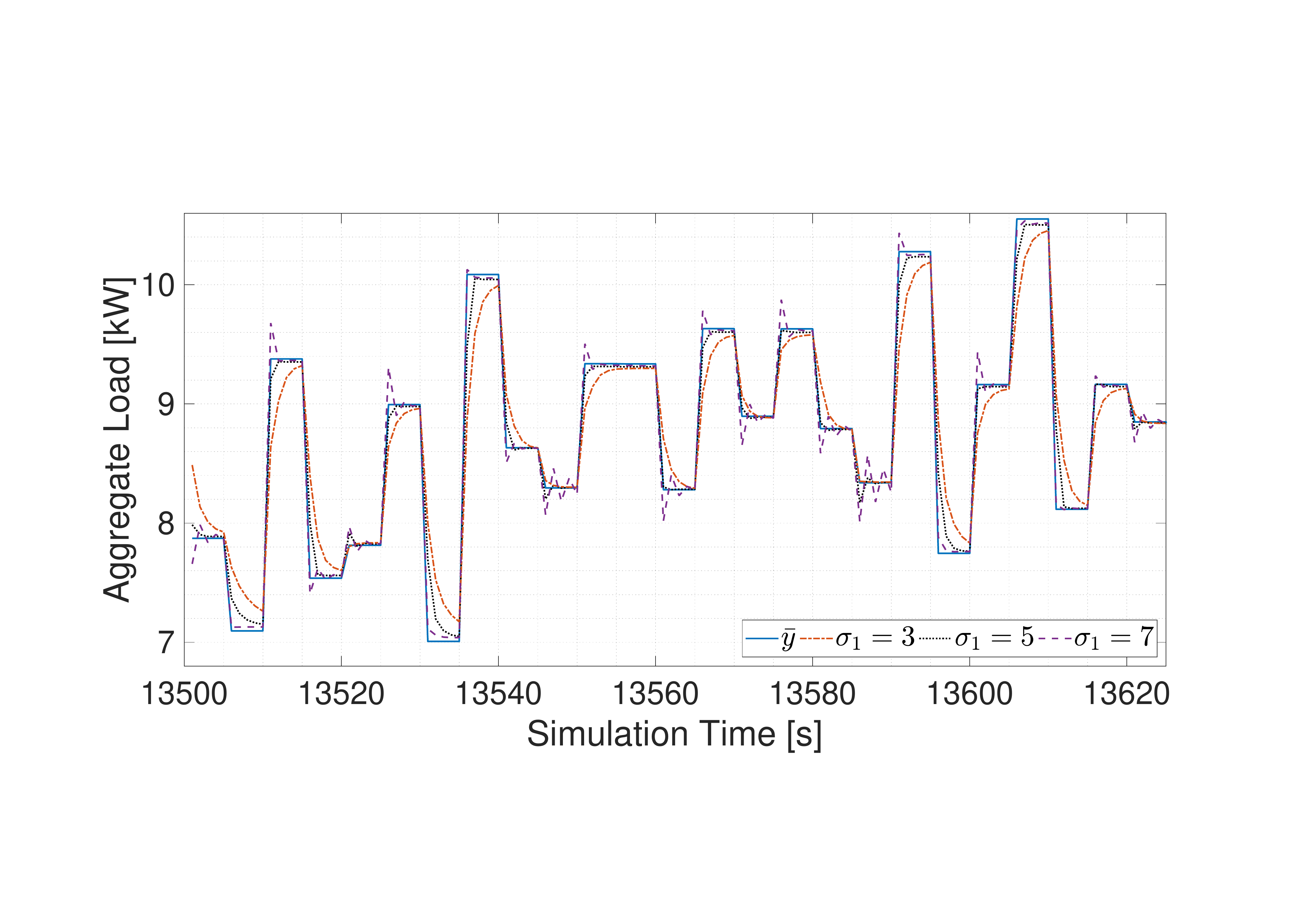} \vspace{-0.1cm}
    \caption{Comparison between different tracking coefficients, $\sigma_1$}
    \label{fig:Sigma1}
    \vspace{-0.5cm}
\end{figure}

As expected, increasing $\sigma_1$ reduces the ancillary service provision NRMSE at the expense of increasing the day-ahead tracking NMAE and most MI estimates. However, a decrease in $\sigma_1$ does not necessarily lead to a decrease in $I_{iid}$ due to a slower convergence rate (see Fig. \ref{fig:Sigma1}). Surprisingly, increasing $\sigma_1$ could also lead to an increase in ancillary service provision MAPE due to tracking overshoots as illustrated in Fig. \ref{fig:Sigma1}. Increasing $\mu_l$ decreases $I_{iid}$, but increases ancillary service provision NRMSE and MAPE, and the average NMAE for tracking the day-ahead schedule.  Changing the sample size $K$ in the objective function has a similar effect to changing $\mu_l$ as it increases the importance of the current control action in estimating the PDF of $(X_l,Y_l)$. However, their effects are not equivalent as reducing $K$ could lead to overfitting the PDF.

\section{Conclusion}
\label{Conclusion}
In this paper, a distributed projected online gradient descent algorithm for providing ancillary services to the grid by aggregating privacy-conscious residential consumers was presented. A balance between the different objectives can be achieved by adjusting their weights. Despite minor performance degradation when compared to an ideal centralised aggregation scheme, the proposed algorithm does not require high-bandwidth communications infrastructure. Moreover, it allows for the preservation of consumer privacy as the actual consumer load does not need to be revealed to the aggregator. 

Future work will focus on the provision of other grid services such as voltage support, incorporating DERs with uncertainty or more complex constraints, and considering grid constraints in the formulation of the optimisation problem.

\bibliographystyle{IEEEtran}
\bibliography{./RTOptTrack}

\end{document}